\documentstyle[12pt,epsf]{article}
\newcommand{\nn}{\nonumber}

\newcommand{\Dsla}{D\hspace{-3.0mm}/}

\newcommand{\pr}{{\mit\Delta}^{-1}}
\newcommand{\W}{W_{\Lambda}}
\newcommand{\G}{\Gamma_{\Lambda}}
\newcommand{\Slash}[1]{\ooalign{\hfil/\hfil\crcr$#1$}}
\newcommand{\psibar}{\bar{\psi}}
\newcommand{\1}{{\bf 1}}
\newcommand{\0}{{\bf 0}}
\newcommand{\C}{{\cal C}}
\newcommand{\F}{{\cal F}}
\renewcommand{\P}{{\cal P}}
\newcommand{\set}{(\psibar\psi)}

\begin{document}
\vspace{-1cm}
\noindent
\begin{flushright}
KANAZAWA-99-13
\end{flushright}
\vspace{20mm}
\begin{center}
{\Large \bf 
Nonperturbative Renormalization Group 
\vspace{2mm}\\
and Renormalizability of Gauged NJL Model
}
\vspace*{15mm}\\
\renewcommand{\thefootnote}{\alph{footnote}}
Ken-Ichi Kubota
\footnote{kubota@hep.s.kanazawa-u.ac.jp}
and Haruhiko Terao 
\footnote{terao@hep.s.kanazawa-u.ac.jp}
\vspace*{3mm}\\
Institute for Theoretical Physics, Kanazawa University\\
Kanazawa 920--1192, Japan
\vspace*{40mm}\\
{\large \bf Abstract}
\end{center}

Non-perturbative renormalizability, or non-triviality, of the gauged 
Nambu-Jona-Lasinio (NJL) model in four dimensions is examined by using
non-perturbative (exact) renormalization group in large $N_c$ limit.
When running of the gauge coupling is asymptotically free and slow 
enough, the two dimensional renormalized trajectory (subspace) spanned 
by the four fermi coupling and the gauge coupling is found to exist, 
which implies renormalizability of the gauged NJL model.
In the case of fixed gauge coupling, renormalizability of the model 
turns out to be guaranteed by the line of the UV fixed points.
We discuss also non-triviality of the gauged NJL model extended to 
include higher dimensional operators and correspondence with the 
gauge-Higgs-Yukawa system.

\newpage
\pagestyle{plain}
\pagenumbering{arabic}
\setcounter{footnote}{0}
\noindent
{\large \bf 1. Introduction}
\vspace*{2mm}

Modern renormalization due primarily to Wilson \cite{WilsonKogut} 
in terms of the scaling of effective Lagrangians offers us quite
intuitive understanding of renormalizability. 
As we scale down to small momenta the effective Lagrangian converges
towards a finite dimensional submanifold, which is called a renormalized
trajectory, in the space of possible Lagrangians.
If this renormalized trajectory is non-trivial, then the theory is
called renormalizable; the effective Lagrangian is given in terms of a
finite number of the renormalized couplings.
Thus renormalizable theories come out as dynamical consequence in the
effective theories in this picture.
This renormalizability in the language of renormalization group
(RG) flows was firstly demonstrated by Polchinski \cite{Polchinski}. 
He gave a proof of perturbative renormalizability of four dimensional 
$\lambda \phi^4$ theories by using the so-called exact RG equation
,or the non-perturbative RG (NPRG) equation.
However the idea of renormalizability itself is non-perturbative and 
expected to be applied beyond perturbation theory, as is mentioned
in his paper \cite{Polchinski}.

There have been known some examples of non-perturbatively 
renormalizable theories. 
These theories have non-trivial continuum limit, or non-trivial
renormalized trajectory, nevertheless they are perturbatively
non-renormalizable.
Four-fermi interaction models \cite{fourfermi} in less than four 
dimensions were shown to be non-perturbatively renormalizable in 1/N
expansion \cite{3dfourfermi,ZinnJustin}. 
Renormalizability of such a model is guaranteed by the presence of 
a non-trivial UV fixed point and the renormalized trajectory comes 
out from the fixed point.
\footnote{Non-linear sigma models in less than four dimensions also 
are known as non-perturbatively renormalizable models with similar 
structure \cite{ZinnJustin}.}
In four dimensions the non-trivial UV fixed point is lost and the 
Nambu-Jona-Lasinio (NJL) model \cite{fourfermi} fails to be 
renormalizable. 

Interestingly, however, it was claimed subsequently by several people 
\cite{KSY,KTY,Krasnikov,HKKN} that the NJL model can be
non-perturbatively renormalizable in four dimensions when the 
fermions are coupled with gauge interaction.
Kondo et. al. investigated the ladder Schwinger-Dyson (SD) equations for
the gauged NJL models and discussed cutoff dependence in the
low energy effective theories \cite{KSY,KTY}.
On the other hand Krasnikov and Harada et. al. found that the 
gauged Higgs-Yukawa theory in four dimensions becomes non-trivial 
under a certain condition by 
analyzing one-loop RG equations in the large $N_c$ limit  
\cite{Krasnikov,HKKN}.
The (gauged) NJL model may be realized as a specific (gauged) Higgs-Yukawa 
theory subject to the so-called compositeness conditions \cite{BHL}.
It was shown \cite{HKKN} that non-triviality of the corresponding
gauge Higgs-Yukawa theory implies renormalizability of the gauged NJL model. 

In this paper we study renormalizability of the gauged NJL model
in four dimensions in the spirit of Polchinski. Namely we examine
existence of the renormalized trajectories spanned by the four-fermi
coupling and the gauge coupling by solving the NPRG equation.
This would be the first application of the NPRG for study of 
non-perturbative renormalizability in four dimensions.
In general we have to truncate the RG equations in some approximation 
scheme for non-perturbative analyses of the solutions. Therefore
it will be a hard problem to give a rigorous proof of non-perturbative
renormalizability. Here we perform the analyses in large N
limit according to the paper \cite{HKKN} in order to make our 
argument definite. 

Dynamics of spontaneous chiral symmetry breaking in the gauged NJL 
model has been minutely investigated mainly by solving the SD 
equations in the (improved) ladder approximation 
\cite{SDfourfermi,KSY,KTY}. Through these analyses the phase structure 
and the anomalous dimensions of this model were clarified.
Recently, however, the method based on the non-perturbative RG equations 
have been invented and been found quite efficient for 
study of dynamical chiral critical behavior \cite{rgqed,rgchiral}. 
In this framework the rather simple beta functions lead directly
to RG flows, phase diagrams, anomalous dimensions, etc. If we truncate
the beta functions so as to include only the ladder type corrections,
then these are found to reproduce the results obtained by solving the
ladder SD equation. However it should be noted that we
may proceed beyond the ladder approximation in this RG approach in a 
systematic way.
In this paper we analyze such non-perturbative RG equations in order 
to see the renormalized trajectories of the gauged NJL models.

The NJL model is understood as a Higgs-Yukawa theory subject to the 
compositeness conditions imposed at some UV scale.
However it has been shown in large N limit that the extended NJL model 
with minimal choice of three independent interaction terms is fully 
equivalent to a Higgs-Yukawa theory \cite{Hasenfratz}.
The Higgs-Yukawa theory in four dimensions is supposed to be trivial
due to presence of the Landau singularity, which may be explicitly shown
in large N limit.
This implies that the non-trivial renormalized trajectory does not 
exist in a rigorous sense.
Hence we do not take this theory non-perturbatively
renormalizable, even though perturbatively renormalizable.
On the other hand asymptotically free gauge interaction can promote the 
Higgs-Yukawa theory to be a non-trivial one \cite{Krasnikov,HKKN}.
Provided the equivalence between the extended NJL models and the
Higgs-Yukawa theories is hold in the presence of gauge interaction, 
the class of non-trivial,
or renormalizable, theories with four-fermi interactions should be
wider than the gauged NJL models. We are going to discuss
renormalizability of these extended theories also. In practice,
the NPRG equations for the gauged NJL models describe evolution of the
extended models as well, since the general possible interactions 
should be equally incorporated in this framework. 

In the next section first let us analyze the exact RG equations for 
large N vector models as a simple example in order to 
discuss renormalized trajectories and non-perturbative renormalizability.
In section 3 we give the flow equations for the gauged NJL model and
show the critical phase structure obtained from these equations.
Section 4 is devoted to our main argument for non-perturbative 
renormalizability of the gauged NJL model. We discuss
renormalizability of the extended gauged NJL models and  
relation to the gauge-Higgs-Yukawa theory in section 5.
\vspace*{2mm}

\noindent
{\large \bf 2. Large N vector model and renormalizability}
\vspace*{2mm}

Before starting the analyses of the gauged NJL model, let us 
briefly discuss the NPRG equations of O(N) vector models
in large N limit
in order to illustrate our strategy and notion of non-perturbative 
renormalizability.
We focus our attention on evolution of the potential part
and study the Wegner-Houghton RG equation with sharp cutoff
\cite{WH}.
For large N vector model in $d$ dimensions the RG equation for
the potential $V(\rho)$ ($\rho=(1/2)(\phi^i)^2$, $i=1,\cdots, N$) 
is given by
\begin{equation}
\frac{dV}{dt}=dV + (2-d)\rho V_{\rho} + a \ln(1+V_{\rho}),
\end{equation}
where we introduced the scale parameter $t=\ln(\Lambda/\mu)$ with
the cutoff scale $\Lambda$ and the renormalization scale $\mu$
and numerical factor $a=\pi^{-d/2}2^{-d}/\Gamma(d/2)$. 
This RG equation turns out to be exact in large N limit
\cite{WH,largeN}.
If we expand the effective potential $V(\rho)$ around it's minimum
\cite{comoving,largeN};
\begin{equation}
V(\rho; \mu) = v_0(\mu) + 
\sum_{n=2}^{\infty}
\frac{\mu^{2(n-2)}}{n!}\lambda_{2n}(\mu)(\rho-\mu^2 \rho_0(\mu))^n,
\end{equation}
the flow equations for the dimensionless couplings are found to be
soluble order by order
\footnote{The Legendre flow equations \cite{legendre} give
the same equations up to the coefficients in the beta functions.}
;
\begin{eqnarray}
\frac{d\rho_{0}}{dt}&=& 2 \rho_0 - a, 
\label{rho} \\
\frac{d\lambda_4}{dt} &=& (4-d)\lambda_4 - a\lambda_4^2, 
\label{lambda4} \\
\frac{d\lambda_6}{dt} &=& (6-2d) \lambda_6 
- a (3\lambda_4\lambda_6 + \lambda_4^3).
\label{lambda6}
\end{eqnarray}
Eq.$(\ref{rho})$ simply gives the critical surface at $\rho_0=a/2$.
The theories in continuum limit must reside in this surface.
It is easy to see existence of the renormalized trajectory in 
$(\lambda_4, \lambda_6)$ space with these simple RG equations.
In four dimension Eq.$(\ref{lambda4})$ is solved as
\begin{equation}
\lambda_4(\Lambda)=\frac{\lambda_4(\mu)}
{1 - a\lambda_4(\mu)\ln(\Lambda/\mu)},
\label{bare4}
\end{equation}
which gives renormalization of $\lambda_4$.
Eq.$(\ref{lambda6})$ also can be easily solved by rewriting it to
\begin{equation}
\frac{d}{dt}\left( \lambda_4^{-3}\lambda_6 \right)
= -2 \lambda_4^{-3}\lambda_6 - a.
\end{equation}
The solution with the renormalized coupling $\lambda_4(\mu)$ 
fixed is found to be
\begin{eqnarray}
\lambda_6(\mu) &=& 
\left( \frac{\mu}{\Lambda}\right)^2
\left( 1 - a \lambda_4(\mu)\ln(\Lambda/\mu)\right)^3
\lambda_6(\Lambda) 
-\frac{a}{2}\lambda_4^3(\mu) 
\left[1-\left(\frac{\mu}{\Lambda}\right)^2\right] \nn \\
&=& F(\lambda_4(\mu)) + 
O\left(\left(\mu/\Lambda\right)^2 \right).
\end{eqnarray}
If we naively make cutoff $\Lambda$ infinity, then $\lambda_6(\mu)$
is determined by a finite function $F$ in terms solely of the 
renormalized coupling
$\lambda_4(\mu)$. This function gives the renormalized trajectory
in this space. 
Similarly all other couplings are determined by $\lambda_4(\mu)$ in
the infinite cutoff limit, which is expected for renormalizable
theory.

However the above argument is not totally correct, since we cannot take 
infinite cutoff limit, or continuum limit, due to the Landau pole 
appearing in Eq.$(\ref{bare4})$. 
Therefore the effective coupling $\lambda_6(\mu)$ suffers from 
ambiguity of order of $(\mu/\Lambda)^2=\exp(-2/a\lambda_4(\mu))$.
\footnote{
Perturbative calculation is blind to such a non-perturbative
effect. Renormalization may be performed by taking infinite cutoff limit 
in perturbative calculation \cite{Polchinski}.} 
Thus we should say that the large N vector model is not non-perturbatively 
renormalizable, though renormalizable perturbatively.
In other words there is no renormalized trajectory in this model.
\vspace*{2mm}

\noindent
{\large \bf 3. Flow equations}
\vspace*{2mm}

There have been proposed several formulations of the NPRG equations, 
which are found to be mutually equivalent.
We employ the formulation called the Legendre flow equation
\cite{legendre}.
This NPRG equation gives the change of the IR cutoff 1PI effective
action under the scale variation leaving the low energy physics unaltered. 

First let us derive the Legendre Flow equation briefly in the case of 
a single scalar field for simplicity.
Appropriate generalization to other kinds of the fields is
straightforward, see Appendix.
We start with the generating functional of IR cutoff connected Green
functions:
\begin{equation}
 e^{\W[J]} 
   = \int {\cal D} \phi \, 
      e^{-\frac12 \phi \cdot \C_{\Lambda}^{-1} \phi
         -S~[\phi] + J \cdot \phi },
\label{IRcutoffGreen} 
\end{equation}
where $\C_{\Lambda}$ is the additive smooth cutoff for the lower
momentum modes than scale $\Lambda$.
From Eq.$(\ref{IRcutoffGreen})$, we can obtain the variation of $\W[J]$
with respect to the cutoff $\Lambda$ as 
\begin{equation}
 \frac{d\W}{d\Lambda}
= -\frac12 \left[ \left( \frac{\delta \W}{\delta J}\right)
                  \cdot \frac{d\C_{\Lambda}^{-1}}{d\Lambda}
	          \left( \frac{\delta \W}{\delta J}\right)
      + \mbox{Tr} \left( \frac{d\C_{\Lambda}^{-1}}{d\Lambda}
                  \cdot \frac{\delta^2 \W}{\delta J \delta J} \right) 
           \right].
\label{floweqforW}
\end{equation}
We define the IR cutoff 1PI effective action by the following
Legendre transformation:
\begin{eqnarray}
 \varphi &=& \frac{\delta \W}{\delta J}, \\
 \G      &=& -\W + J \cdot \varphi 
             - \frac12 \varphi \cdot 
               \C_{\Lambda}^{-1} \varphi.
\label{Legendretr}
\end{eqnarray}
From Eq.$(\ref{floweqforW})$ and Eq.$(\ref{Legendretr})$, we obtain
\begin{eqnarray}
 \frac{d\G}{d\Lambda} 
&=& \frac12 \mbox{Tr} \left[ \frac{d\C_{\Lambda}^{-1}}{d\Lambda}
    \left( \C_{\Lambda}^{-1} + 
     \frac{\delta^2 \G}{\delta \varphi \delta \varphi}
    \right)^{-1}      \right],
\label{floweq}\\
&=& \frac12 \mbox{Tr} \left[ \frac{d\C_{\Lambda}^{-1}}{d\Lambda}
    \left( \pr_{>} + \left.
     \frac{\delta^2 \G}{\delta \varphi \delta \varphi}
     \right|_{
     \raisebox{5pt}[5pt][5pt]{
     \begin{tabular}{@{$\!$}l} 
      {\tiny{interaction}} \\[-10pt] 
      {\tiny{part}}  
     \end{tabular}}}
  \right)^{-1}      \right],
\label{originalfloweq}
\end{eqnarray}
where we define IR cutoff propagator $\mit{\Delta}_{>}$ with
$\mit{\Delta}$ of the effective action $\G$ by
\begin{equation}
 \pr_{>} = \C_{\Lambda}^{-1} + \pr .
\end{equation}

The right hand side of Eq.$(\ref{originalfloweq})$  may  be regarded
as sum of one-loop graphs by expanding the inverse matrix with respect
to the cutoff propagator $\mit{\Delta}_{>}$. 
Thus the beta functions of the couplings are easily evaluated. 
In general, however, the flow equations for the infinitely many
couplings are intertwined each other. 
Therefore it is necessary to truncate the series of the couplings by
some approximation scheme so as to solve the equations.

The chiral critical behavior of the gauged NJL models has been 
examined by applying the Legendre flow equations 
\cite{rgqed,rgchiral}.
The RG flow of the four-fermi coupling of NJL type plays an 
essential role to determine the phase. 
In the large N (and ladder) approximation, which is adopted in the
later considerations the flow equations are reduced to be rather simple.
In Fig.1 and Fig.2 the RG flows projected on the space of the
four-fermi coupling and the gauge coupling are shown in the case
that the gauge coupling is fixed and is asymptotically free
respectively 
\cite{rgchiral}. 
These should be compared with those obtained by
solving the SD equations 
\cite{SDfourfermi,KSY,KTY}.
In Ref.\cite{rgqed} the gauge independent flow equations for all the 
chirally invariant four-fermi interactions have been derived in a 
certain approximation scheme better than the large N and ladder.

\epsfxsize=0.5\textwidth
\leavevmode
\epsffile{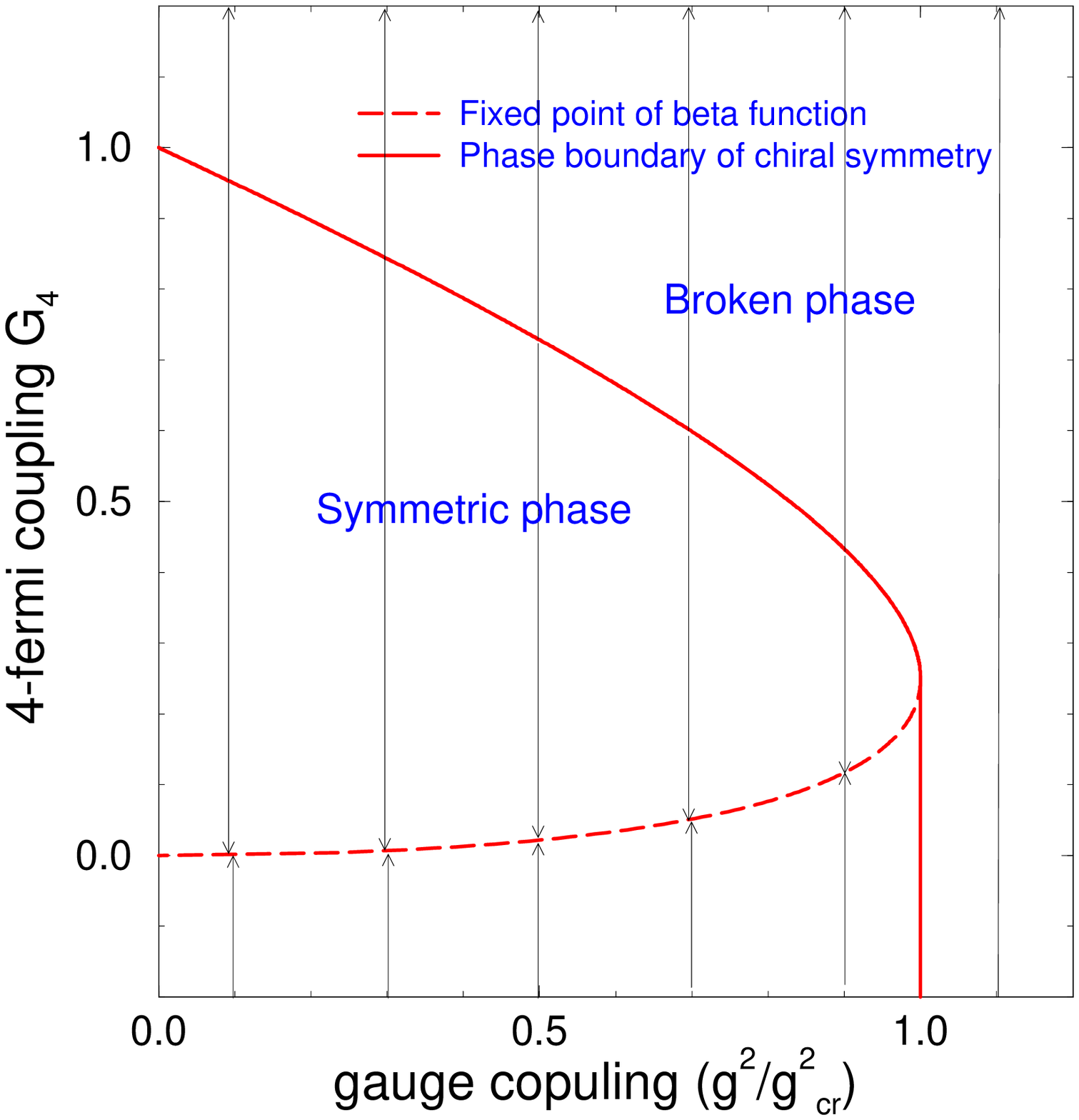}
\epsfxsize=0.5\textwidth
\leavevmode
\epsffile{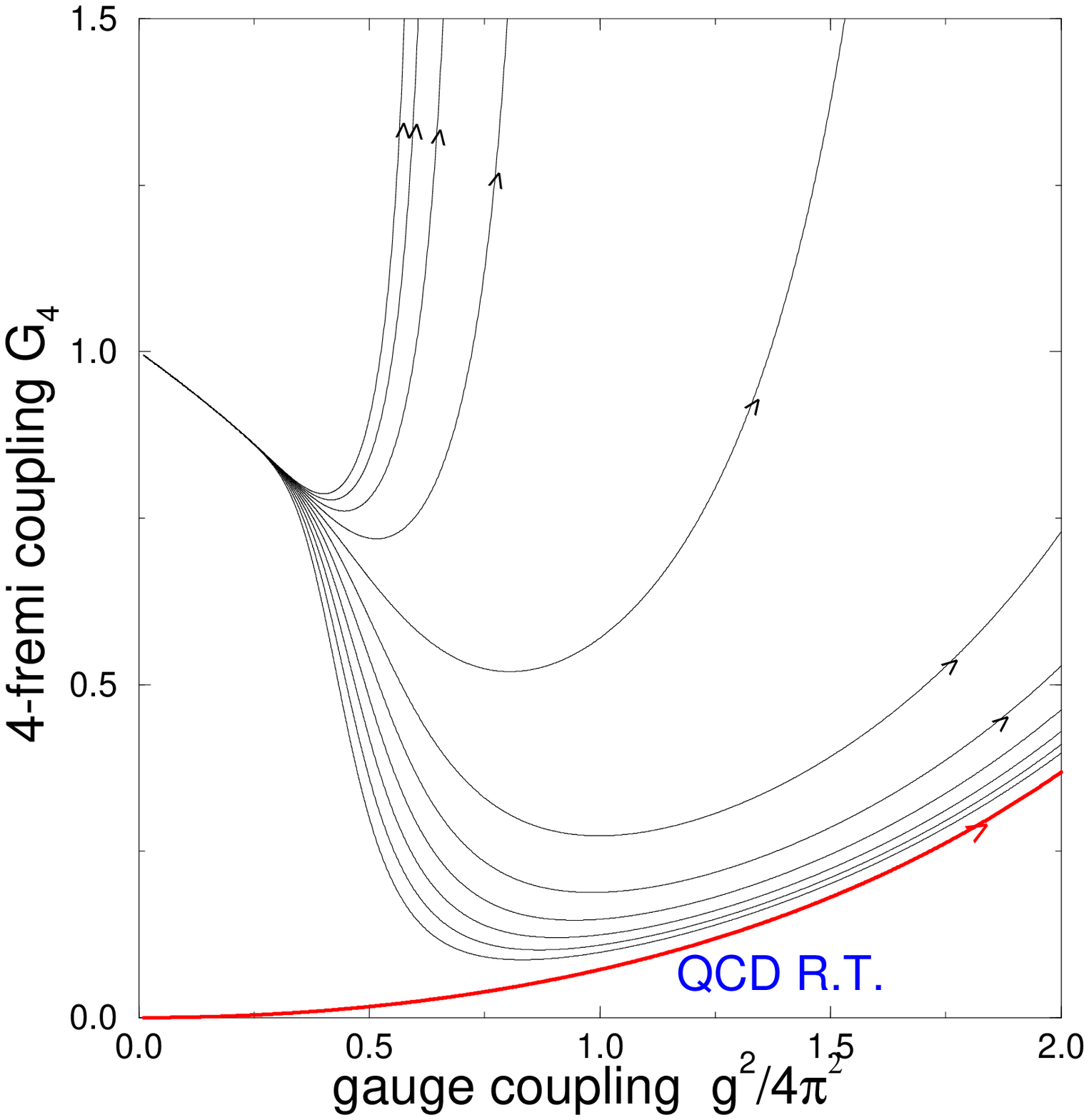}
\begin{center}
\parbox[t]{75mm}
{Fig.1: RG flows and the chiral phase boundary of 
the gauged NJL model with
fixed gauge coupling shown in the $(g^2,G_4)$ plane.}
\hspace*{5mm}
\parbox[t]{75mm}
{Fig.2: RG flows of the gauged NJL model with 
asymptotically free gauge coupling shown in the $(g^2,G_4)$ plane. 
The bold line represent the renormalized trajectory of QCD projected on 
this plane. }
\end{center}
\vspace{0.5cm}

It is seen from Fig.1 that there are two phases in the
gauged NJL model with fixed gauge coupling. The upper (lower) region
is found to be chirally broken (unbroken) phase \cite{rgchiral}.
As long as the four-fermi coupling is concerned, the critical surface
of the phase boundary appears to be a line of UV fixed points.
The theories possessing an UV fixed point are renormalizable, since
there should exist the renormalized trajectories subject to the
fixed point.
Therefore it is supposed that the gauged NJL model is also
renormalizable.
However it is necessary to examine the RG flows of other couplings
than the four-fermi as well in order to confirm the UV fixed points
and the renormalizability in this model.

When we take account of the asymptotically free evolution of the 
gauge coupling, the phase boundary is washed out as is seen in
Fig.2.
QCD has a non-trivial continuum limit given by the renormalized
trajectory subject to the Gaussian fixed point, which is shown by the 
bold line in Fig.2. However other flows of the gauged NJL models 
do not seem to
converge on the renormalized trajectory of QCD. The four-fermi
couplings of all these flows start
at the critical point of the NJL model in the UV limit.
Therefore it may be supposed
that these flows are on another renormalized trajectory subject to
the non-trivial UV fixed point.
If so, the gauged NJL models offers non-perturbatively renormalizable
theories different from QCD.

In the next section we are going to examine the renormalizability
of these theories by studying behavior of the RG flows.
The clue observation is to see existence of the non-trivial
UV fixed point and the renormalized trajectories.
For this purpose it is enough to clarify the RG flows in the region
of weak gauge coupling. Hence we are going to evaluate the beta functions
in the lowest non-trivial order of gauge coupling in
the followings.

Now we are going to present the flow equations for the $SU(N_c)$ gauged NJL
model in the leading order of $1/N_c$ expansion, which will be
investigated in the next section.
Let us introduce $N_f$ flavors of fermions in the fundamental
representation $\psi_i (i=1, \cdots, N_f)$, and assume $N_f$ is order
of $N_c$ so as to tune the gauge beta function in large $N_c$ limit.
Also we assume that only $n$ ($\sim O(1)$) species among them have 
four-fermi interactions according to Harada et.al.\cite{HKKN}.
Here we restrict the four-fermi interaction to the Gross-Neveu type 
\cite{fourfermi} for simplicity.
Namely our starting bare Lagrangian at scale $\Lambda$ is given by
\begin{equation}
{\cal L}_{\Lambda}=\frac14 \mbox{Tr} F^2_{\mu \nu}
+ \sum_{i=1}^{N_f} \psibar^i \Dsla \psi^i 
- \frac{1}{2 \Lambda^2}G_4(\Lambda)
\left(\sum_{i=1}^{n}\bar{\psi^i}\psi^i\right)^2.
\end{equation}

\par
In general quantum corrections generate all the operators consistent 
with the symmetry of the system in the effective Lagrangian. 
Here, however, we may treat gauge coupling perturbatively, since we are
interested in the high energy asymptotic region $(g^2 \ll 1)$.
In the large $N_c$ limit the effective Lagrangian at scale $\mu$
is found to be given by the following form:
\begin{eqnarray}
{\cal L}_{\mu}&=&\frac14 \mbox{Tr} F^2_{\mu \nu}
+ \sum_{i=1}^{N_f} \psibar^i \Dsla \psi^i 
- \frac{1}{2 \mu^2}G_4(\mu)
\left(\sum_{i=1}^{n}\bar{\psi^i}\psi^i\right)^2 \nonumber\\
& &
- \frac{1}{2 \mu^4}G'_4(\mu)
\left(\sum_{i=1}^{n}\bar{\psi}^i\psi^i\right)\partial^2
\left(\sum_{j=1}^{n}\bar{\psi}^j\psi^j\right)
- \frac{1}{4 \mu^6}G_8(\mu)
\left(\sum_{i=1}^{n}\bar{\psi}^i\psi^i\right)^4
+ \cdots.
\label{gNJL}
\end{eqnarray}
No other interactions such as $(\bar{\psi}\gamma_{\mu}\psi)^2$ are
generated in the large $N_c$ limit.
\par
The Legendre flow equations for the gauged NJL model may be obtained
by substituting the above Lagrangian into Eq.(\ref{floweq}).
However it will be found that the RG flows for the couplings
$(g^2, G_4, G_4', G_8)$ are relevant to examine renormalizability.
Therefore let us show the flow equations only for these couplings,
which are found to be
\begin{eqnarray}
\frac{dg^2}{dt} &=& b g^4,
\label{g}\\
\frac{dG_4}{dt} &=& -2 G_4 + w G_4^2 + c g^2 G_4, 
\label{G4}\\
\frac{dG'_4}{dt} &=& -4 G'_4 + 2w G_4 G'_4 + aG_4^2 + c g^2 G'_4 + e
g^2 G_4,
\label{G4'}\\
\frac{dG_8}{dt} &=& -8 G_8 + 4w G_4G_8 + d G_4^4 + 2c g^2 G_8 + f g^2
G_4^3,
\label{G8}
\end{eqnarray}
where the scale parameter $t$ denotes $\ln (\Lambda/\mu)$.
These beta functions may be obtained by evaluating the one-loop
diagrams schematically shown in Fig.3.  
It should be noted that higher dimensional multi-fermi operators 
do not affect evolution of the lower dimensional ones. 
Hence we may solve these equations order by order without truncation.  
The derivation of Eq.(\ref{g})-Eq.(\ref{G8}) and the explicit values of the
coefficients are given in Appendix. 

\begin{figure}[hbt]
\begin{flushleft}
 \leavevmode
 \hspace*{15mm}
 \epsfxsize=8cm
 \epsfbox{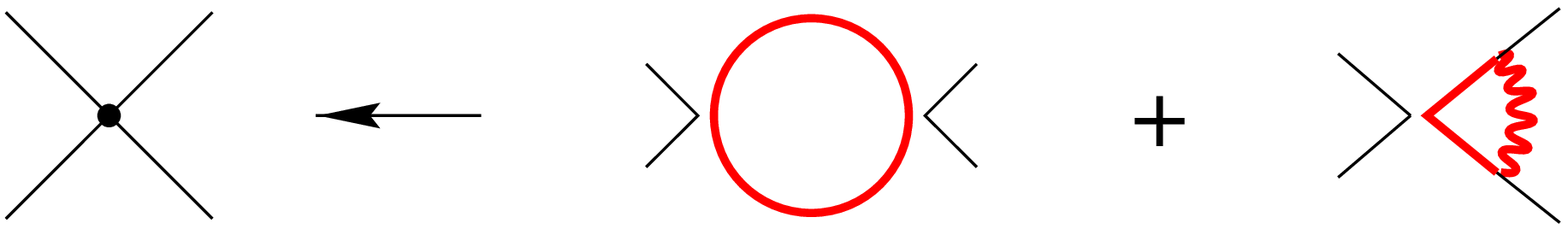}
\vspace*{5mm}
 \hspace*{15mm}
 \epsfxsize=13cm
 \epsfbox{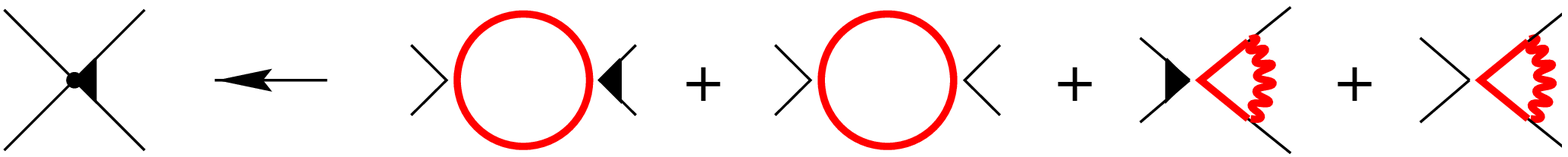}
\vspace*{5mm}
 \hspace*{15mm}
 \epsfxsize=13cm
 \epsfbox{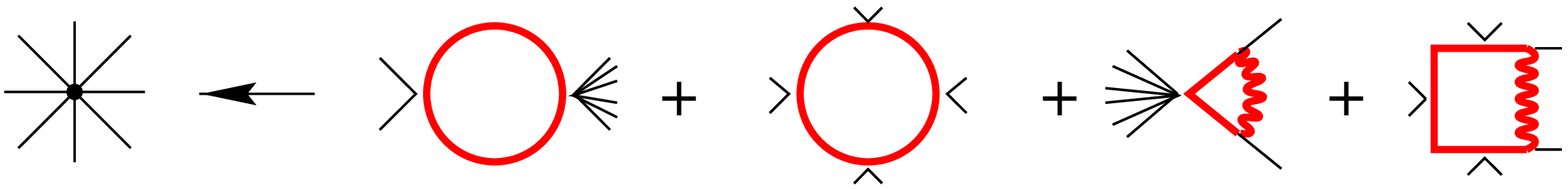}
\end{flushleft}
\begin{center}
\parbox[t]{160mm}{
Fig.3: The quantum corrections for the effective couplings 
$(g^2, G_4, G_4', G_8)$ given by the Legendre flow equation
are shown by one-loop diagrams schematically.
}
\end{center}
\end{figure}  
\vspace*{2mm}

\noindent
{\large \bf 4. Renormalizability of the gauged NJL model}
\vspace*{2mm}

Now let us analyze the solution of the RG flow equations given by
Eq.$(\ref{g})$-$(\ref{G8})$ to see renormalizability of the gauged NJL model.
For this purpose we fix the renormalized couplings $g^2(\mu)$ and
$G_4(\mu)$ at some scale $\mu \ll \Lambda$.
Also the bare couplings other than $g(\Lambda)$ and $G_4(\Lambda)$
are set to zero.
Note that the renormalized trajectory of QCD is given by $G_4(\mu)=0$, 
since we ignored $O(g^4)$ corrections in the flow equations.

First we give $(g^2(\Lambda),G_4(\Lambda))$ explicitly in terms of the
renormalized couplings. Solution of Eq.$(\ref{g})$ is the well-known one;
\begin{equation}
g^2(\Lambda) = \frac{g^2(\mu)}{1+bg^2(\mu)\ln(\Lambda/\mu)},
\label{solution:g}
\end{equation} 
which shows asymptotically free running of the gauge coupling.
Eq.$(\ref{G4})$ may be solved by using the trick illustrated in 
section 2.
Assuming that $G_4$ is positive and non-vanishing, which is the case
of our present interest, Eq.$(\ref{G4})$ can be 
rewritten as
\begin{equation}
\frac{d}{dt}\left( g^{2c/b} \frac{1}{G_4} \right)
= 2g^{2c/b} \frac{1}{G_4} - w g^{2c/b}.
\end{equation}
The solution of this equation is easily found to be 
\begin{equation}
G_4(\Lambda) = G_4(\mu) \left(\frac{\Lambda}{\mu}\right)^2
\frac{( 1+bg^2(\mu)\ln(\Lambda/\mu) )^{-c/b}}
{1 + w G_4(\mu) \int_{\mu}^{\Lambda}\frac{d\Lambda'}{\Lambda'}
\left(\frac{\Lambda'}{\mu}\right)^2
( 1+bg^2(\mu)\ln(\Lambda'/\mu) )^{-c/b}},
\label{solution:G4}
\end{equation}
from which we can see cutoff dependence of the bare coupling $G_4(\Lambda)$.
In the infinite cutoff limit, $G_4(\Lambda)$ is found to take a finite 
value; 
\begin{eqnarray}
\lim_{\Lambda \rightarrow \infty} G_4(\Lambda) 
&=& \lim_{\Lambda \rightarrow \infty}
\frac{( 1+bg^2(\mu)\ln(\Lambda/\mu) )^{-c/b}}
{\frac{1}{G_4(\mu)} \left(\frac{\mu}{\Lambda}\right)^2
 + \frac{w}{2} ( 1+bg^2(\mu)\ln(\Lambda/\mu) )^{-c/b}} \nn \\
&=& \frac{2}{w} .
\end{eqnarray}
Thus it is seen that all the flows with $G_4(\mu)>0$ are found to 
approach to the NJL critical coupling $G_4^*=2/w$ in UV limit.
It is seen from Eq.$(\ref{solution:G4})$ also that if we perform 
perturbative calculation, the bare coupling diverges more than 
quadratically as cutoff goes to infinity.
Such divergence is tamed away by non-perturbative sum of radiative
corrections. 

Next we examine the RG flow equations for $G'_4$ and $G_8$; 
Eq.$(\ref{G4'})$ and Eq.$(\ref{G8})$. 
By noting that these equations are also rewritten into
\begin{eqnarray}
\frac{d}{dt}\left(g^{2c/b}\frac{G'_4}{G_4^2}\right)
&=& g^{2c/b}\left(a + e g^2 \frac{1}{G_4}\right), \\
\frac{d}{dt}\left(g^{4c/b}\frac{G_8}{G_4^4}\right)
&=& g^{4c/b}\left(d + f g^2 \frac{1}{G_4}\right),
\end{eqnarray}
we may easily derive the explicit form of the solutions in terms of
$g^2(\Lambda)$ and $G_4(\Lambda)$ given in 
Eq.$(\ref{solution:g})$ and Eq.$(\ref{solution:G4})$.
They are found to be
\begin{eqnarray}
G'_4(\mu)&=& 
G'_4(\Lambda)\left(\frac{g^2(\Lambda)}{g^2(\mu)}\right)^{c/b} 
\left(\frac{G_4(\Lambda)}{G_4(\mu)}\right)^{-2} \nn \\
& & + G_4(\mu)^2 \int_{\mu}^{\Lambda} \frac{d\Lambda'}{\Lambda'}
\left(\frac{g^2(\Lambda')}{g^2(\mu)}\right)^{c/b} 
\left( a + e \frac{g^2(\Lambda')}{G_4(\Lambda')}\right), 
\label{solution:G4'}\\
G_8(\mu)&=& 
G_8(\Lambda)\left(\frac{g^2(\Lambda)}{g^2(\mu)}\right)^{2c/b} 
\left(\frac{G_4(\Lambda)}{G_4(\mu)}\right)^{-4} \nn \\
& & + G_4(\mu)^4 \int_{\mu}^{\Lambda} \frac{d\Lambda'}{\Lambda'}
\left(\frac{g^2(\Lambda')}{g^2(\mu)}\right)^{2c/b} 
\left( a + f \frac{g^2(\Lambda')}{G_4(\Lambda')}\right),
\label{solution:G8}
\end{eqnarray}
where dependence on the bare couplings $G_4'(\Lambda)$ and
$G_8(\Lambda)$ are also given for the later discussions.
In order for the gauged NJL model to be renormalizable, there have to
exist the renormalized trajectory determined by the renormalized
couplings $g^2(\mu)$ and $G_4(\mu)$. This means that the integrations 
in Eq.$(\ref{solution:G4'})$ and in Eq.$(\ref{solution:G8})$ should
converge as $\Lambda$ goes to infinity.
To see this let us evaluate the first derivatives with respect to
cutoff at $\Lambda \gg \mu$;
\begin{eqnarray}
\Lambda\frac{dG'_4(\mu)}{d\Lambda} 
&\sim& a G_4(\mu)^2 (1 + bg^2(\mu)\ln(\Lambda/\mu) )^{-c/b} \nn \\
& & + e G_4(\mu)^2 \frac{g^2(\mu)}{G_4^*}
(1 + bg^2(\mu)\ln(\Lambda/\mu) )^{-c/b-1}, 
\label{derivative:G4'}\\
\Lambda\frac{dG_8(\mu)}{d\Lambda} 
&\sim& d G_4(\mu)^4 (1 + bg^2(\mu)\ln(\Lambda/\mu) )^{-2c/b} \nn \\
& & + f G_4(\mu)^2 \frac{g^2(\mu)}{G_4^*}
(1 + bg^2(\mu)\ln(\Lambda/\mu) )^{-2c/b-1}. 
\label{derivative:G8}
\end{eqnarray}
From these equations the conditions for convergence are found to be
\begin{eqnarray}
& & b < c \hspace{12mm}\mbox{for finite $G'_4$}, \\
& & b < 2c \hspace{10mm}\mbox{for finite $G_8$}.
\end{eqnarray}
Note that the convergence is quite slow due to logarithmic damping,
which is clear difference from power damping behavior of irrelevant 
couplings around a fixed point. 

So far we have been concerned with evolution of only the three effective
couplings $(G_4, G'_4, G_8)$ for fermionic operators with less
dimensions. In principle, however, we need to examine cutoff
dependence of all other effective couplings to see renormalizability.
Actually it would be seen that the cutoff dependence of other
couplings decays with power behavior. Here let us show convergence of
$G_{12}$ coupling of $(\bar{\psi}\psi)^6$ only. 
The beta function of $G_{12}$ in the present approximation scheme is
given by
\begin{equation}
\frac{dG_{12}}{dt} = -14 G_{12} + 6w G_4G_{12} + 3w G_8^2 - 6d G_4^3
G_8 + f G_4^6 + 3c g^2 G_{12} + 9 e g^2 G_4^2G_8 + h g^2 G_4^5.
\end{equation}
By performing similar manipulation we may find the solution for
$G_{12}(\mu)$ as follows
\begin{eqnarray}
G_{12}(\mu) 
&=& \left(\frac{\mu}{\Lambda}\right)^2
\left(\frac{g^2(\Lambda)}{g^2(\mu)}\right)^{3c/b}
\left(\frac{G_4(\Lambda)}{G_4(\mu)}\right)^{-6}G_{12}(\Lambda) \\ \nonumber
& & + G_4(\mu)^6 \int_{\mu}^{\Lambda} \frac{d\Lambda'}{\Lambda'}
\left(\frac{\mu}{\Lambda'}\right)^2
\left(\frac{g^2(\Lambda')}{g^2(\mu)}\right)^{3c/b} 
\left(3w\frac{G_8}{G_4^6} - 6d\frac{G_8}{G_4^3} + f +
9e \frac{G_8}{G_4^4} + h \frac{g^2}{G_4}
\right)(\Lambda').
\end{eqnarray}
It is seen that the cutoff dependence of $G_{12}$ is quadratically
suppressed and $G_{12}(\mu)$ converges rapidly to a finite value as 
$\Lambda$ goes to infinity. 
Cutoff independence of other higher couplings may be shown also 
inductively. 

Thus the renormalized trajectory (subspace) given by the flow
equations is found to be determined with $(G_4, G'_4, G_8)$ and 
gauge coupling in general.
In the case of $b<c$ this subspace is reduced to the two dimensional
one spanned by $g^2$ and $G_4$. The aspect of convergence 
to the renormalized trajectory will be seen from the
flow diagrams projected on $(g^2,G'_4)$ plane and $(g^2,G_8)$ plane
as shown in Fig.4 for $c=3b$.
The renormalized trajectory is shown by the bold lines.
The renormalized trajectory goes to infinity of the coupling space 
rapidly as the gauge coupling approaches to 0. 
In Fig.5 the RG flows for $c=0.1b$ are also shown. 

\begin{figure}[hbt]

\epsfxsize=0.5\textwidth
\leavevmode
\epsffile{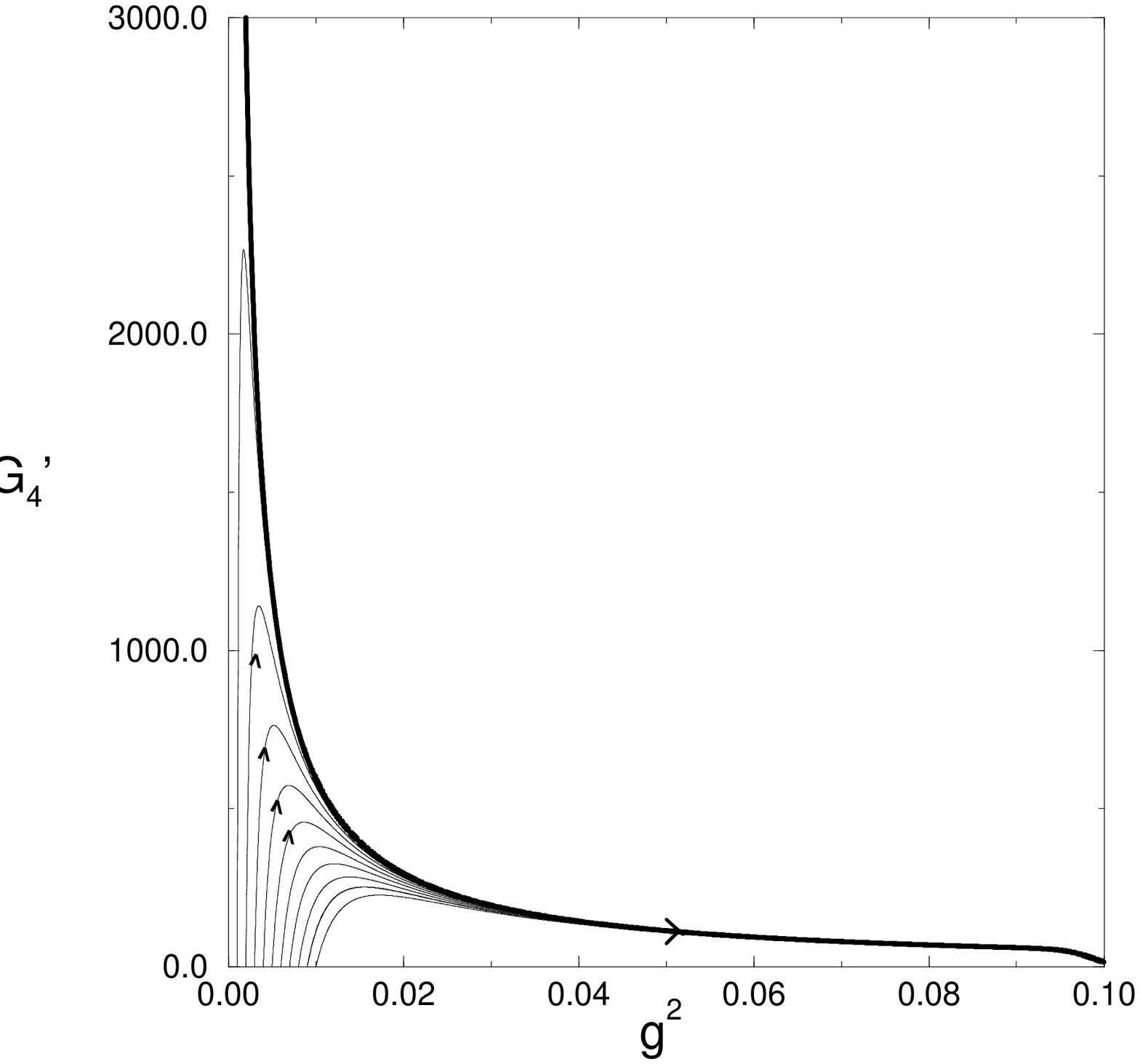}
\epsfxsize=0.5\textwidth
\leavevmode
\epsffile{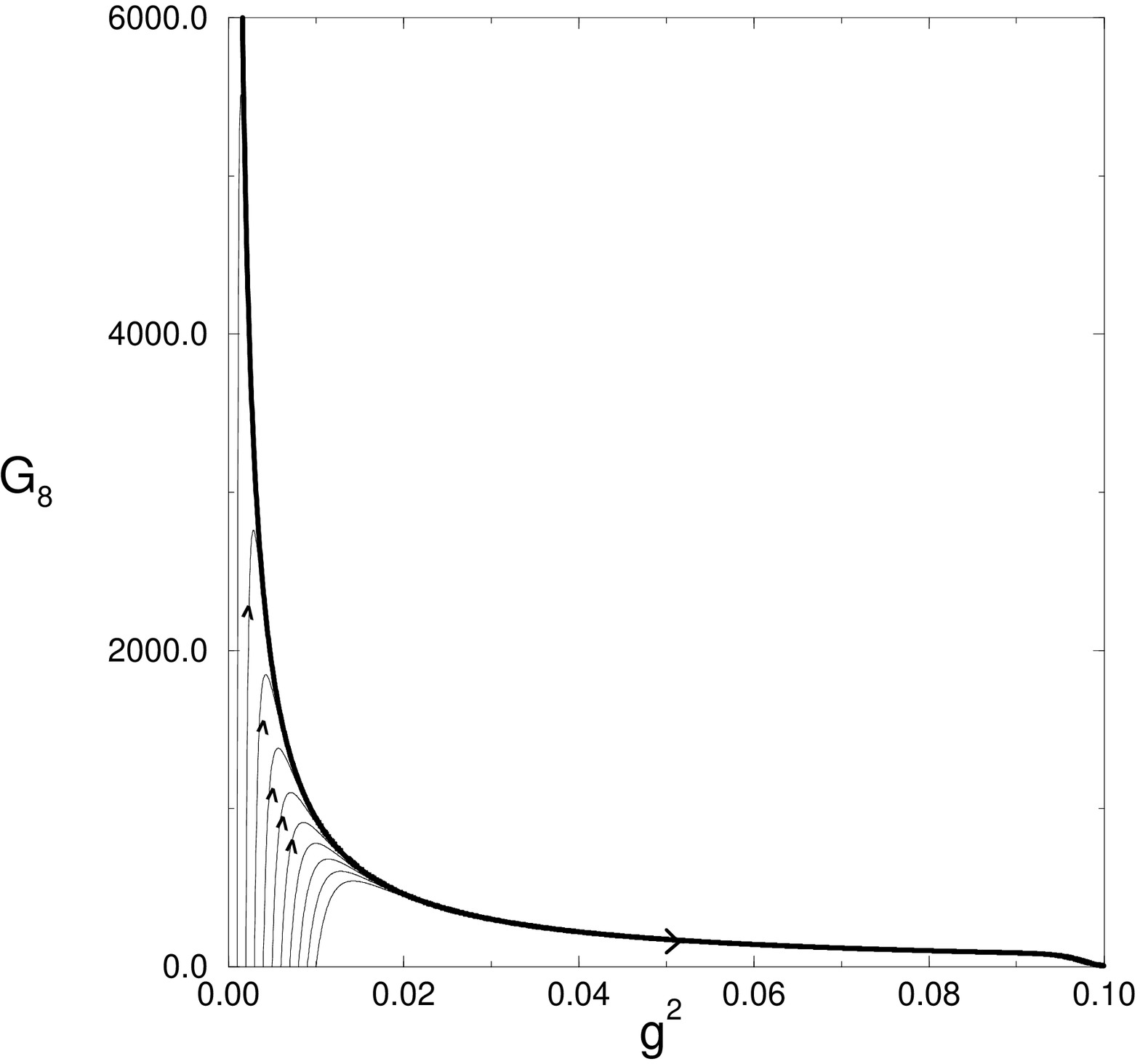}

\begin{center}
\parbox[t]{130mm}
{Fig.4: The RG flows of the gauged NJL model with $c=3b$
projected on $(g^2,G'_4)$ plane (left) and on $(g^2,G_8)$ plane (right). 
The imposed renormalization conditions are 
$g^2(\mu)=0.1 \times \frac{c}{b}$ and $G_4(\mu)=1/w$.
The bold lines show the renormalized trajectory.
}
\end{center}

\end{figure}
\begin{figure}[hbt]

\epsfxsize=0.5\textwidth
\leavevmode
\epsffile{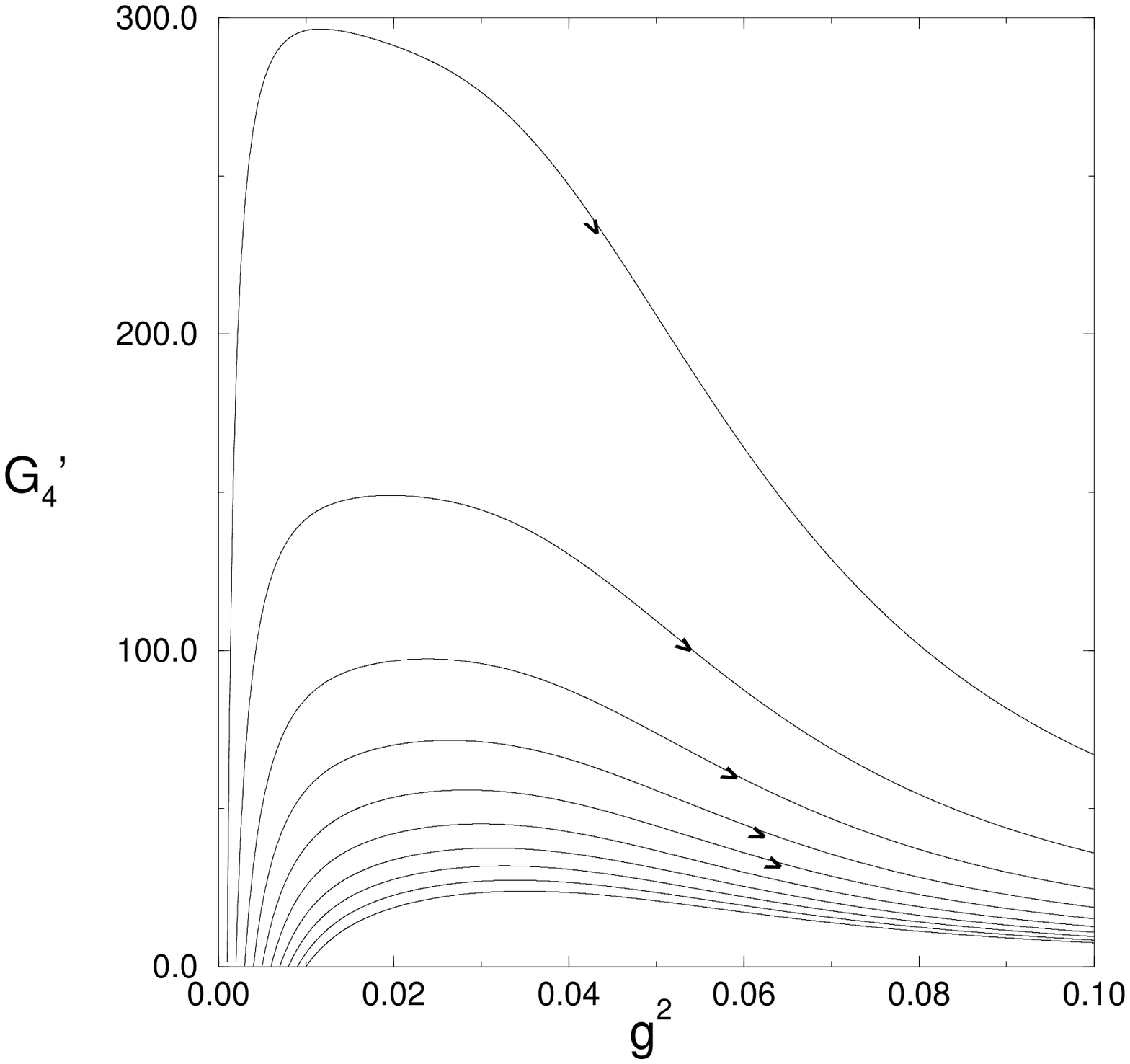}
\epsfxsize=0.5\textwidth
\leavevmode
\epsffile{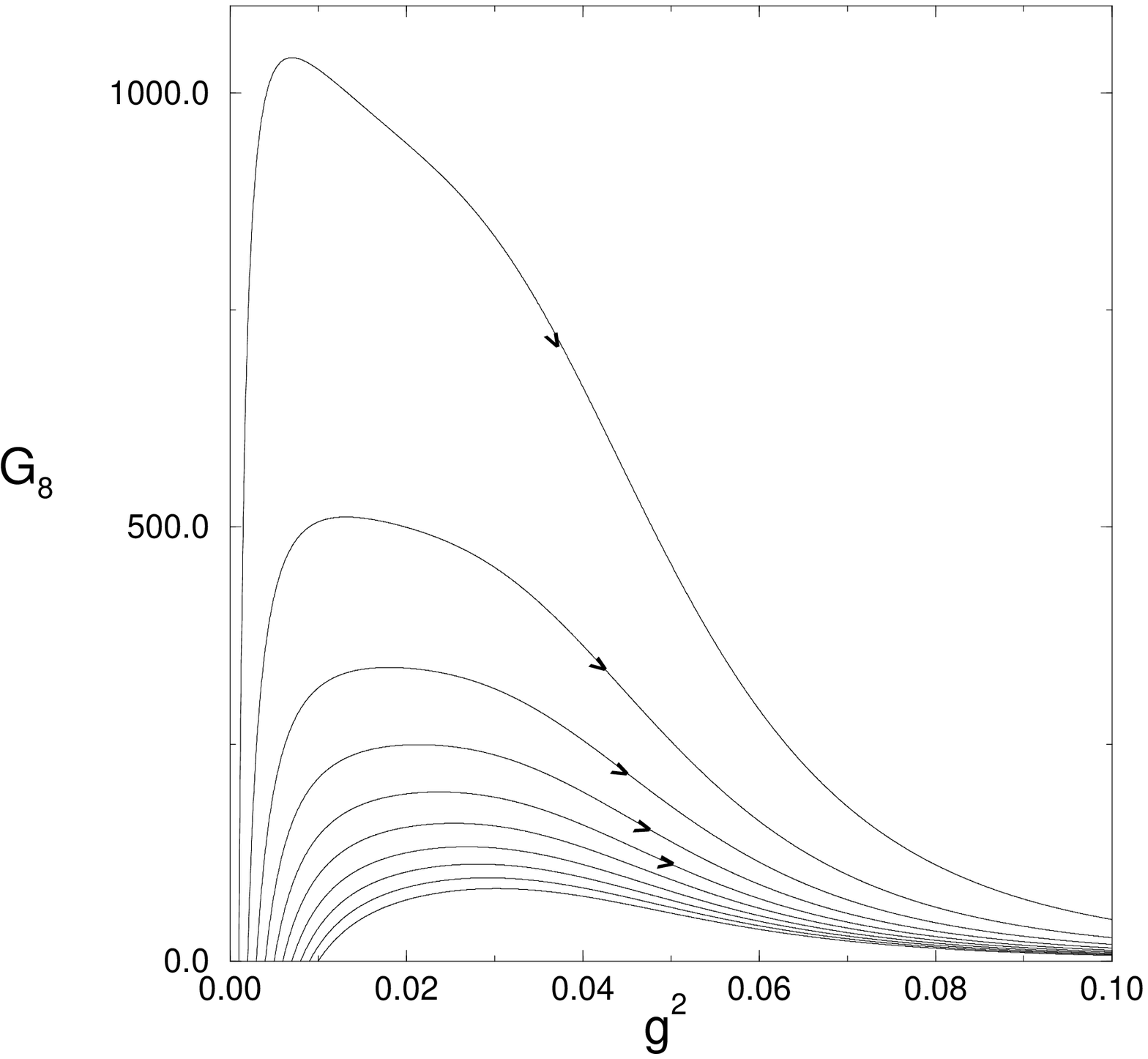}

\begin{center}
\parbox[t]{130mm}
{Fig.5: The RG flows of the gauged NJL model with $c=0.1b$
projected on $(g^2,G'_4)$ plane (left) and on $(g^2,G_8)$ plane (right). 
The imposed renormalization conditions are 
$g^2(\mu)=0.1 \times \frac{c}{b}$ and $G_4(\mu)=1/w$.
}
\end{center}

\end{figure}

Thus the gauged NJL model turns out to be non-perturbatively 
renormalizable when the coefficients of the beta functions satisfy 
$b < c$ \cite{HKKN,KSY,Krasnikov}.
However this does not means that there are no renormalizable theories
unless $b<c$. If we allow renormalization of couplings 
$G'_4$ and/or $G_8$, then we could obtain other classes of renormalized
theories with non-trivial continuum limit. We are going to discuss
this point in the next section.

For the fixed gauge coupling the line of fixed points appears and aspect of
the flow diagram is rather different from the asymptotically free case. 
\footnote{Rather the flow diagram is similar to the NJL model in $d <
4$, which is known to be renormalizable in 1/N expansion.}
The solutions of the flow equations with the fixed gauge coupling $g^2$ 
may be obtained by taking 
limit of $b \rightarrow 0+$ in Eq.$(\ref{solution:G4})$, 
Eq.$(\ref{solution:G4'})$ and Eq.$(\ref{solution:G8})$ with noting the 
relation
\begin{equation}
\lim_{b \rightarrow 0+}
\left(\frac{g^2(\Lambda)}{g^2(\mu)}\right)^{c/b} 
= \lim_{b \rightarrow 0+}
(1 + bg^2\ln(\Lambda/\mu) )^{-c/b}
=\left(\Lambda/\mu\right)^{-cg^2}.
\label{fixedlimit}
\end{equation}
From Eq.$(\ref{solution:G4})$ $G_4$ is solved as
\begin{equation}
\left(\frac{1}{G_4(\mu)}-\frac{1}{G_4^*(g)}\right)
=\left(\frac{g^2(\Lambda)}{g^2(\mu)}\right)^{2-cg^2}
\left(\frac{1}{G_4(\Lambda)}-\frac{1}{G_4^*(g)}\right),
\end{equation}
where $G_4^*(g)=(2-cg^2)/w$ is the fixed line coupling.
This solution implies that $1/G_4$ is relevant with dimension
$2-cg^2$ around the fixed line. 
By using the relation Eq.$(\ref{fixedlimit})$  
in Eq.$(\ref{derivative:G4'})$ and in Eq.$(\ref{derivative:G8})$ 
it is seen that cutoff dependence of $G'_4$ and $G_8$ in turn 
decays with power law as long as the gauge coupling is non-vanishing;
\begin{eqnarray}
\Lambda\frac{dG'_4(\mu)}{d\Lambda}&=& 
O\left(\left(\mu/\Lambda\right)^{cg^2}\right), \\ 
\Lambda\frac{dG_8(\mu)}{d\Lambda}&=& 
O\left(\left(\mu/\Lambda\right)^{2cg^2}\right). 
\end{eqnarray}
The powers of convergence correspond to the dimensions of
these irrelevant couplings, dim$[G'_4]=-cg^2$ and dim$[G_8]=-2cg^2$.
Thus the continuum limit of the gauged NJL model 
with fixed gauge coupling is described by renormalized trajectory
coming out from the fixed point and the theory is clearly 
renormalizable.
\par
\vspace*{2mm}

\noindent
{\large \bf 5. Extension of the gauged NJL model}
\vspace*{2mm}

In the previous section we saw that the renormalized trajectory
of the gauged NJL model is determined by the renormalized gauge
coupling and the four fermi coupling, when $c>b$. 
In other cases, $G'_4$ and/or $G_8$ diverge in the infinite cutoff
limit, however other couplings for the higher dimensional operators
remain irrelevant. 
Therefore we may suppose that the renormalized trajectory of the theory
with $c\leq b$, if it exists, is enlarged so as to be determined by 
the renormalized couplings $(g^2, G_4, G'_4, G_8)$.
This implies that there can be a larger class of renormalizable
theories, if we allow to renormalize the couplings $G'_4$ and $G_8$ 
as well.
However the extended NJL models with any renormalized couplings 
$(g^2, G_4, G'_4, G_8)$ will be seen not always renormalizable, 
even though all the other higher dimensional operators are irrelevant.
Specially the NJL model without gauge interaction cannot be made
renormalizable by extension. In this section we discuss what kinds of
the extended NJL models can be renormalizable.

Indeed this issue is related with non-triviality of the gauge-Higgs-Yukawa 
theories \cite{Krasnikov,HKKN};
\begin{equation}
{\cal L}^{\mbox{\scriptsize gHY}}=
\frac{1}{4}\mbox{Tr}F_{\mu\nu}^2
+ \sum_{i=1}^{N_f}\bar{\psi}^i\Dsla\psi^i + 
\frac{1}{2}(\partial_{\mu}\phi)^2 + y\phi  \sum_{i=1}^n \bar{\psi}^i\psi^i
+ \frac{1}{2}m_{\phi}^2 \phi^2 + \frac{1}{8} \lambda \phi^4. 
\label{gHY}
\end{equation}
The gauged NJL models discussed in the previous section are known to
be equivalent to the gauge-Higgs-Yukawa theories subject to the
compositeness condition at some composite scale $\Lambda$ \cite{BHL};
\begin{equation}
\frac{m_{\phi}^2(\Lambda)}{y^2(\Lambda)}=\frac{1}{G_4}(\Lambda),~~~
\frac{1}{y^2(\Lambda)}=0,~~~
\frac{\lambda(\Lambda)}{y^4(\Lambda)}=0.
\label{comp}
\end{equation}
However Hasenfratz et. al. claimed that the extended NJL models with three
independent fermionic interactions are equivalent to the Higgs-Yukawa
theories without such condition in large N limit \cite{Hasenfratz}. 
The Green functions in those theories are made to be the same through 
identification of the renormalized coupling constants. 

Such matching between the effective couplings in these theories
can be explicitly seen also by the exact RG equations.
By similar calculation done in section 3 the RG equations for the 
dimensionless couplings of the gauge-Higgs-Yukawa theory are found to be
\footnote{The form of these flow equations are same as those examined 
in Ref.\cite{HKKN}. The beta function for the scalar mass of this form
is obtained in mass dependent renormalization scheme.} 
\begin{eqnarray}
\frac{dm_{\phi}^2}{dt} &=& 2m_{\phi}^2 - w y^2 -a y^2 m_{\phi}^2, 
\label{mass}\\
\frac{dy}{dt} &=& -\frac{a}{2}y^3 + \frac{c}{2}g^2 y, 
\label{y}\\
\frac{d\lambda}{dt} &=& -2a y^2 \lambda + dy^4,
\label{lambda}
\end{eqnarray}
where we truncated the effective action to the perturbatively 
renormalizable one given by Eq.$(\ref{gHY})$. 
Actually these equations are found to be transformed into the RG 
equations for the gauged NJL model given by
Eq.$(\ref{G4})$-$(\ref{G8})$ through the following non-linear relations;
\begin{equation}
\frac{1}{G_4} = \frac{m_{\phi}^2}{y^2},~~~
\frac{G'_4}{G_4^2} = \frac{1}{y^2},~~~
\frac{G_8}{G_4^4} = \frac{\lambda}{y^4}.
\label{matching}
\end{equation}
However the last terms in Eq.$(\ref{G4'})$  and in Eq.$(\ref{G8})$
are not reproduce these relations though their contribution to the
effective couplings are irrelevant for the arguments of
renormalizability.
These terms are supposed to be related with radiative
corrections to the higher dimensional operators, 
$\partial^2\phi \bar{\psi}\psi$ and $\phi^3\bar{\psi}\psi$ in the
gauge-Higgs-Yukawa theory discarded in the above calculation. 
Here, however, let us just neglect these subleading contributions.
It should be  also noted that the compositeness conditions given by
Eq.(\ref{comp}) are nothing but the requirement of vanishing $G'_4$
and $G_8$ at scale $\Lambda$.

Thus there exists a mapping between the RG flows of the extended
gauged NJL model and of the gauge-Higgs-Yukawa theory. 
Especially the renormalized trajectory of the gauged NJL model
discussed in the previous section corresponds to the solution of the
so-called coupling constant reduction \cite{Kubo} in the
gauge-Higgs-Yukawa theory.
In the theories along this flow $y(\mu)$ and $\lambda(\mu)$ are
completely determined by the gauge coupling $g(\mu)$ and both are
asymptotically free.
This renormalized trajectory in the gauge-Higgs-Yukawa theory, 
which is infrared attractive by it's definition, is also known as
the Pendleton-Ross (PR) fixed point \cite{PR}.  

Harada et. al. investigated triviality and stability bound for the
gauge-Higgs-Yukawa theory by solving the RG equations Eq.$(\ref{y})$  
and Eq.$(\ref{lambda})$  with Eq.$(\ref{g})$
\cite{HKKN}. 
They found that the allowed region of the renormalized couplings 
$(y, \lambda)$ survive 
on a line connecting the Gaussian fixed point and the PR fixed point 
in the infinite cutoff limit as long as $c>b$, as is shown in Fig.3 in 
ref.\cite{HKKN}.
Now these theories are parametrized by renormalized couplings 
$g^2$, $m_{\phi}$ and $y$.
Therefore this result implies that there are non-trivial or 
non-perturbatively
renormalizable theories with three renormalized couplings $g^2$, $G_4$
and $G'_4$ among the extended gauged NJL models when $c>b$.
The coupling correspondence given by Eq.$(\ref{matching})$  tells us 
the conditions which these non-trivial theories should satisfy at 
the cutoff scale are
\footnote{The stability bound of the scalar potential
$\lambda(\Lambda) \geq 0$ is mapped to $ G_8(\Lambda) \geq 0$.
The triviality bound implies that 
$G_8(\Lambda) < \infty$ should be imposed.
However, this bound in the gauge-Higgs-Yukawa theory may
be obtained by assuming triviality of pure $\lambda\phi^4$ theory 
in four dimensions, which is not seen in the large $N_c$ 
analysis.
At present it is unclear to us how this triviality bound may be
imposed to the extended gauged NJL model.
}
\begin{equation}
0 \leq G'_4(\Lambda), ~~~~0 \leq G_8(\Lambda) < \infty.
\end{equation}
Indeed it is seen from Eq.$(\ref{solution:G4'})$  and
Eq.$(\ref{solution:G8})$ that continuum limit with 
these boundary conditions may be taken
only if $c>b$. The renormalized $G'_4$ off the renormalized trajectory
of the gauged NJL model is achieved by tuning the bare coupling 
$G'_4(\Lambda)$ properly. 
It is supposed that these conditions should be imposed on the extended
NJL model so as to satisfy unitarity and stability for collective 
excitations of $\bar{\psi}\psi$ bound state.

\vspace{5mm}
\noindent
{\bf Acknowledgment}
\vspace*{2mm}

The authors thank to K-I.~Aoki and M.~Tomoyose for valuable
discussions
and also to H.~Onoda for collaboration in the early
stage.

\vspace*{5mm}
\renewcommand{\theequation}
{A.\arabic{equation}}
\setcounter{equation}{0}
\noindent
{\large \bf Appendix}
\vspace*{2mm}

In this appendix we derive the flow equations given by 
Eq.(\ref{g})-Eq.(\ref{G8}) with explicit coefficients.
To this end, we first advance Eq.(\ref{floweq}) to a more appropriate
form for our analysis.
Let us define the ``propagator'' and the ``vertex'' as
\begin{eqnarray}
 \P^{-1} &=& {\bf 1} + \pr \C_{\Lambda} ,\\
 \F &=& \left.
     \frac{\delta^2 \G}{\delta \varphi \delta \varphi}
     \right|_{
     \raisebox{5pt}[5pt][5pt]{
     \begin{tabular}{@{$\!$}l} 
       {\tiny{interaction}} \\[-10pt] 
       {\tiny{part}}  
     \end{tabular}}}\C_{\Lambda} ,
\label{vertex}
\end{eqnarray}
then Eq.(\ref{floweq}) becomes
\begin{equation}
 \frac{d\G}{d\Lambda} 
= -\frac12 \mbox{Tr} \left[ \C_{\Lambda}^{-1} \frac{d\C_{\Lambda}}{d\Lambda} 
  \left( \P^{-1} + \F   \right)^{-1}      \right].
\label{additiveeq}
\end{equation}
Next we rewrite the Eq.(\ref{additiveeq}) in terms of the
dimensionless quantities by canonical scaling.
Also we perform wave function renormalization so that the kinetic term 
is always normalized.
By taking account of wave function renormalization it is convenient to 
introduce the following form of the cutoff function.
\begin{equation}
\C_{\Lambda}^{-1}(p,q) = Z C^{-1}(q^2/\Lambda^2) \Lambda^2
(2\pi)^4\delta^4(p+q).
\end{equation}
where the function $C(q^2/\Lambda^2)$ cuts off 
lower momentum modes smoothly.
$Z$ is wave function renormalization.
Thus the Legendre flow equation is given by 
\begin{eqnarray}
    \frac{d}{dt}\Gamma[\varphi;t] 
&=& \left[ 
     D - \int \frac{d^D q}{(2\pi)^D} 
     \varphi(q) 
     \left( d_{\varphi}+\frac{\eta_{\varphi}}{2}+q_{\mu}
     \frac{\partial '}{\partial q_{\mu}} \right)
     \frac{\delta}{\delta \varphi (q)}
    \right] \Gamma[\varphi;t] \nonumber \\
& & \quad +
    \int \frac{d^D q}{(2\pi)^D} 
    \left[\left\{
          \left( 
        \frac{q^2}{C} \frac{\partial C}{\partial q^2} + 1
          \right)
          -\frac{\eta_{\varphi}}{2}
          \right\}
          \left( \P^{-1} + \F \right)^{-1} {}_{q,q}
    \right].
\label{LegendreFlow}
\end{eqnarray}
The first line of Eq.(\ref{LegendreFlow}), where $\eta_{\varphi}$
denotes the anomalous dimension defined by 
$\eta_{\varphi} = -\Lambda d(\ln Z)/d\Lambda $, 
represents nothing but the canonical scaling of the cutoff effective
action.

By expanding the inverse matrix in the second line as 
\begin{equation}
 \left( \P^{-1} + \F \right)^{-1} 
= \P - \P\F\P + \P\F\P\F\P - \cdots.
\end{equation}
we may obtain the beta function for each coupling by comparing the
coefficients of each operator on the both sides of
Eq.(\ref{LegendreFlow}).
Each term, which is regarded as a one-loop Feynman graph, can be
evaluated by giving $\C, \P,$ and $\F$.

For the gauged NJL model of our present concern we need treat fermions as well as bosons.
Then the elements appearing in the Legendre flow equation, $\C, \P,$ and $\F$ 
become matrices in $\{ A_{\mu}, \psi, \psibar \}$ space.
Tr is replaced with sTr.
As we notice that
$\C_{\Lambda}(p,q) = \C_{\Lambda}(q) (2\pi)^4 \delta^4 (p+q)$ and
$\P(p,q) = \P(q) (2\pi)^4 \delta^4 (p+q)$, 
in the large $N_c$ limit these matrices found to be given as follows;
\begin{eqnarray}
\C(q) &=& 
C(q^2)
\left(
 \begin{array}{c|c}
   \1_4 \otimes \1_{N_c} & \0 \\ \cline{1-2}\\[-0.4cm]
  \0 &
  \left(
  \begin{array}{cc}
   0 & -i \Slash{q}^T \\
   -i \Slash{q} & 0 
  \end{array}
  \right)   \otimes \1_{N_f \times N_c}
 \end{array}
\right),\\
\P(q) &=& 
\left(
 \begin{array}{c|c}
   G_T \left(\delta_{\mu \nu} - \frac{q_{\mu} q_{\nu}}{q^2}\right) 
   \otimes \1_{N_c} & \0 \\ \cline{1-2}
  \0 &
  \left(
  \begin{array}{cc}
	S\1_4 & 0 \\
 	0     & S\1_4
  \end{array}
  \right)   \otimes \1_{N_f \times N_c}
 \end{array}
\right),\\
\F(p,q) &=& 
C(q^2)
\left(
 \begin{array}{c|c}
  \0 & g \psi_i^T \gamma_{\nu}^T \Slash{q}^T T_a^T  \quad 
      -g \psibar_i^T \gamma_{\nu} \Slash{q} T_a \\ \cline{1-2}\\[-0.4cm]
   \left.
   \begin{array}{c}
     i g \gamma_{\mu}^T T_a^T \psibar_j \\
    -i g \gamma_{\mu}   T_a   \psi_j 
   \end{array}
   \right. &
   \left(
   \begin{array}{cc}
    g \Slash{A}_a^T \Slash{q}^T T_a^T & 0 \\
    0 & g \Slash{A}_a \Slash{q} T_a
   \end{array}
   \right) \otimes \1_{N_f}
 \end{array}
\right) \\
& & + C(q^2)
\left(
 \begin{array}{c|c}
  \0 & \0 \\ \cline{1-2}\\[-0.4cm]
  \0 &
  \left(
   \begin{array}{cc}
    -i g G \Slash{q}^T \otimes (\1_{n}\oplus\0_{N_f-n})& 0 \\
    0 & i g G \Slash{q}\otimes (\1_{n}\oplus\0_{N_f-n})
   \end{array}
  \right) \otimes \1_{N_c} \nonumber
 \end{array}
\right), 
\end{eqnarray}
where we used the Landau gauge propagator for the gauge field.
Note that there is no anomalous dimension for the fermion field in the
Landau gauge, and that anomalous dimension for the gauge field do not
contribute to the beta functions in the calculation ${\cal O}(g^2)$.
We introduce also the following functions in the above equations.
\begin{eqnarray}
 G_T(q^2) &=& \frac1{1+q^2 C(q^2)},\\
 S(q^2)   &=& \frac1{1+q^2 C(q^2)},
\end{eqnarray}
\begin{equation}
 G = G_4 \set - G_4' (-p+q)^2 \set + G_8 \set^3 + \cdots .
\end{equation}
We substitute these matrices for the Legendre flow equation, then 
calculating the part of the radiative corrections of the Legendre flow
equation return in calculating the Feynman graph given in Fig.3.
It is straightforward to obtain the results 
Eq.(\ref{G4})-Eq.(\ref{G8})
by using the technique of the derivative expansion \cite{Morris}.
The coefficients in the beta functions are given by the following form;
\begin{eqnarray} 
w &=& +16 N_c n A \int_0^{\infty} q^3 \,dq \;q^2 \xi C^2 S^3 \\
\label{w}
a &=& -8 N_c n A \int_0^{\infty} q^3 \,dq \;\xi C S^2
     [3 q^2 (CS)' + q^4 (CS)''] \\ 
c &=& +12 C_2(R) A \int_0^{\infty} q^3 \,dq \;\xi 
       q^2 C^3 S^2 G_T (G_T + 2S) \\
d &=& -32 N_c n A \int_0^{\infty} q^3 \,dq \;\xi q^4 C^4 S^5 \\
e &=& +3 C_2(R) A \int_0^{\infty} q^3 \,dq \;\xi 
     \left\{
      C^2 S G_T^2 [3 q^2 (CS)' + q^4 (CS)'' ] \right.\nonumber\\
  & & \quad \left.
     +C S^2       [3 q^2 (C^2 S G_T)' + q^4 (C^2 S G_T)'' ]
     +C^2 S^2 G_T [3 q^2 (CS)' + q^4 (CS)'' ]
     \right\} \\
f &=& -24 C_2(R) A \int_0^{\infty} q^3 \,dq \;\xi
       q^4 C^5 S^4 G_T (G_T + 4 S)
\label{f}
\end{eqnarray}
where
$\xi = (q^2 C'/C +1)$ 
and 
$A =\frac{1}{8\pi^2}$.
The prime denotes the derivative with respect to $q^2$.

Here we treat the gauge coupling perturbatively, therefore
the coefficient of the gauge beta function is just the one loop one;
\begin{equation}
 b = \frac{11 N_c - 4 T(R) N_f}{3}.
\label{b}
\end{equation}


\end{document}